# X-ray Luminosity Function and Total Luminosity of LMXB in Early Type Galaxies

Dong-Woo Kim and Giuseppina Fabbiano

Harvard-Smithsonian Center for Astrophysics
60 Garden Street, Cambridge, MA 02138

Apr. 19, 2004

Abstract

We have derived bias-corrected X-ray luminosity functions (XLFs) of sources detected in a uniformly selected sample of 14 E and S0 galaxies observed with *Chandra* ACIS-S3. The entire sample yields 985 point-like X-ray sources, with typical detection of 30-140 sources per galaxy. After correcting for incompleteness, the individual XLFs are statistically consistent with a single power-law of a (differential) XLF slope β = 1.8 - 2.2 (with a typical error of 0.2-0.3). A break at or near $L_{X,Eddington}$, as reported in the literature for some of these galaxies, is not required in any case. Given the uniform XLF shape, we have generated a combined, higher statistics, XLF, representative of X-ray sources in elliptical galaxies. Although the combined XLF is marginally consistent with a single power-law (with β = 2.1 ± 0.1), a broken power-law gives an improved fit. The best-fit slope is β = 1.8 ± 0.2 in the low-luminosity range $L_X$ = a few x $10^{37}$ – 5 x $10^{38}$ ergs s$^{-1}$. At higher luminosities, the slope is steeper, β = 2.8 ± 0.6. The break luminosity is 5 ± 1.6 x $10^{38}$ ergs s$^{-1}$ (with an error at 90%), which may be consistent with the Eddington luminosity of neutron stars with the largest possible mass (3 M☉), of *He*-enriched neutron star binaries, or low-mass stellar mass black holes. If the change in XLF slope at high luminosities is real, and does not mask a step in the XLF, our result would imply a different population of high luminosity sources, instead of a beaming effect. This high luminosity portion of the XLF must reflect the mass function of black holes in these galaxies. We note that this high luminosity population does not resemble that of the ultra luminous X-ray sources (ULX) detected in star-forming galaxies, where no break in the XLF is present, and the XLF is much flatter than in the older stellar system we are studying here. We use our results to derive the integrated X-ray luminosity of accreting low-mass X-ray binaries (LMXBs) in each sample galaxy. We confirm that the total X-ray luminosity of LMXBs is correlated with the optical and more tightly with the near-IR luminosities, but in both cases the scatter exceeds that expected from measurement errors. We find that the scatter in $L_X$(LMXB)/$L_K$ is marginally correlated with the specific frequency of globular clusters.

*Subject headings*: galaxies: elliptical and lenticular, cD – X-rays: binaries – X-rays: galaxies

# 1. Introduction

With its sub-arcsecond spatial resolution and unprecedented sensitivity (Weisskopf et al. 2000), *Chandra* has provided incontrovertible proof of the existence of populations of LMXBs in all E and S0 galaxies (e.g., Sarazin et al. 2001; Angelini et al. 2001; Kim and Fabbiano 2003, hereafter **KF03**). Prior to *Chandra*, there was good but indirect evidence of the existence of these LMXBs (see Fabbiano 1989). This evidence included both statistical considerations, and the spectral properties of the integrated X-ray emission. The former were based on the observation that the lower envelope of the X-ray / optical luminosity distribution of E and S0 galaxies is consistent with the integrated output of their stellar population and X-ray binary component, inferred from the extrapolation of the properties of the bulge of M31 (e.g., Trinchieri & Fabbiano 1985; Eskridge et al 1995). The latter evidence was the detection of a hard spectral component, typical of LMXBs, in both the *Einstein* (Kim et al. 1992), and more strikingly in the wide-beam *ASCA* spectra (Matsushita et al. 1994) of these galaxies; this hard emission, however, could also have stemmed from low luminosity or absorbed AGNs (e.g., Allen, Di Matteo & Fabian 2000). The *Chandra* images reveal populations of point-like sources in all E and S0 galaxies. The *Chandra* ACIS spectra of these sources are hard, as would be expected from LMXBs (see review in Fabbiano & White 2003).

The detection of X-ray source populations in galaxies with *Chandra* has created a veritable industry in the derivation of the X-ray luminosity functions (XLFs) of these populations (see Fabbiano & White 2003 and refs. therein). Comparison of XLFs of different star-forming galaxies provides empirical evidence of differences in their star formation history (e.g. Gilfanov et al 2003). Comparison with X-ray binary evolution models tailored to the different stellar populations shows that predictions and consistency checks for both the shape and the normalization of XLFs are possible with theoretical X-ray binary modeling (Belczynski et al 2003). These results show that the XLFs of sources in a given galaxy (or galaxian region) reflect the formation, evolution, and physical properties of the X-ray source population.

The XLFs of the E and S0 galaxies observed with *Chandra* are generally steeper than those of star-forming galaxies, and are fitted with power-law (or broken power-laws) with reported cumulative slopes ranging from 1.0 to 1.8. Breaks have been reported both at luminosities of ~2 x $10^{38}$ ergs s$^{-1}$, near the Eddington luminosity of neutron stars (e.g., Sarazin et al. 2000; Blanton et al. 2001; Kundu et al. 2002), and at higher luminosities near $10^{39}$ ergs s$^{-1}$ (Jeltema et al. 2003). While the former break may be related to a transition between neutron star and black-hole binaries as the main contributors to a given luminosity range (Sarazin et al 2000), the high luminosity break could be produced by an aging 'younger' component from binaries formed in rejuvenation (merging) episodes (Jeltema et al. 2003).

These results (and conclusions), however, may be affected by incompleteness effects, as demonstrated by **KF03** in the case of NGC 1316. Low luminosity sources may be missed in the inner region of galaxies, because of the increased background levels from the more intense diffuse emission (see also Zezas & Fabbiano 2002). At larger radii, the detection sensitivity is affected by the widening of the *Chandra* PSF. In NGC 1316, these effects resulted in an apparent break at ~2 x $10^{38}$ ergs s$^{-1}$ that disappeared, producing a straight power law, once the incompleteness corrections were applied. This result affects both our understanding of the X-ray population of NGC 1316, and also the estimate of the integrated contribution of fainter LMXBs to the total unresolved X-ray emission. As discussed in **KF03**, the latter would affect both the measure of the metal abundances of the hot ISM, and measurements that rely on the knowledge of the properties of the hot ISM, such as the determination of the binding mass of the galaxy assuming hydrostatic equilibrium.

The purpose of the work reported in this paper is to explore how typical the results of **KF03** are. We extended our study to a well-defined sample of 14 E and S0 galaxies, all observed with the ACIS-S3 CCD chip. All the data used in this work were retrieved from the public *Chandra* archive, and some of them have been used in publications by the original *Chandra* investigators. For all these galaxies, we have derived bias corrected XLFs, with incompleteness corrections (following **KF03)** applied in all cases.

This paper is organized as follows. In Section 2, we describe our sample selection and data reduction technique. In Section 3, we first derive and correct for incompleteness the XLF of each galaxy in the sample, concluding that are all consistent with each other within statistics; we then derive a combined XLF for the entire sample, to increase our sensitivity. In Section 4, we use these results to calculate the total X-ray luminosity of LMXBs and its ratio to the galaxy optical luminosity, using both detected and undetected LMXBs. In Section 5, we discuss the features of the composite XLF, and we compare our derived LMXB integrated luminosities with both galaxian optical (stellar) luminosities and globular cluster specific frequency, to set constraints on LMXBs formation and evolution scenarios. Finally, we summarize our conclusions in Section 6.

## 2. Sample selection and data reduction

To build XLFs for a sizeable sample of early type galaxies, we have extracted data from the *Chandra* public archive (http://asc.harvard.edu/cda/), according to the following selection criteria: (1) we have selected only early type galaxies with morphological type T < 0 (taken from **RC3**), which (2) were the targets of *Chandra* pointings, resulting in the best possible angular resolution and point-source detection sensitivity; (3) to minimize instrumental effects, we have used only data taken with the ACIS-S3 (back-illuminated) CCD chip, which is better calibrated than the front-illuminated chips, and is also more sensitive at low energies; (4) to ensure low enough detection thresholds, we have further restricted the sample to include only observations with exposures times longer than 20 ksec, after background flare screening; finally, (5) we require that there are more than 30 detected points sources, detected within the $D_{25}$ ellipse (25-th magnitude isophote, taken from **RC3**) of a given galaxy, so as to have a reasonable determination of the XLF. Our selection method ensures that our sample consists of uniform data and that the instrumental effects can be corrected with best-known calibration data. The resulting sample consists of 14 early type galaxies. Their basic properties are listed in Table 1, together with the published papers on these data. We take the distance from the surface brightness fluctuation results by Tonry et al. (2001). For galaxies in the Virgo and Fornax clusters, we adopt the group distance and for the remaining galaxies, we use individual measurements. For all galaxies except three, the data were acquired after the ACIS temperature reached −120°C (which is the best-calibrated state so far). The gain correction is less accurate for NGC 1399, NGC 4636, and NGC 4697, hence spectral information may suffer from larger systematic uncertainties for these galaxies. However, the broad-band photometry needed for this study would not be significantly affected.

We have uniformly reduced the ACIS data with XPIPE (Kim et al. 2004a), a suite of software specifically developed for the *Chandra* Multi-wavelength Project (ChaMP). XPIPE takes the CXC pipeline Level 2 data products and then applies additional corrections (e.g., gain correction, removing bad pixels/columns) and screening (e.g., removing background flares). For details, we refer to Kim et al. (2004a). To detect X-ray sources, we used the CIAO (http://cxc.harvard.edu/ciao) **wavdetect** tool (Freeman & Kashyap 2002). As discussed in **KF03** and Kim et al. (2004a), the performance and limitations of **wavdetect** are well understood and calibrated by extensive simulations.

We determined source fluxes in the 0.3-8 keV band by applying an energy conversion factor (ECF) to the detected source count rate. The ECFs had to be tailored to each observation, because the quantum efficiency (QE) of ACIS varies with time (for the QE degradation, see CXC Memo on Jul. 29, 2002; http://cxc.harvard.edu/cal/Acis/Cal_prods/qeDeg/index.html) and because the Galactic value of $N_H$ varies from one pointing to another. To calculate the ECF, we assume a power-law emission model of $\Gamma_{ph}$ = 1.7 with absorption by Galactic $N_H$ determined for each observation (Stark et al. 1992), and use the CIAO **Sherpa** package (http://cxc.harvard.edu/sherap), in conjunction with **corrarf,** a tool that corrects the instrumental response files to take into account the variation of the QE (available from httpcxc.harvard.edu/cal/Acis/Cal_prods/qeDeg/corrarf.tar.gz). We find that the ECFs vary by ~25% within our sample.

**3. X-ray Luminosity Functions**

To construct the XLF of the X-ray source population of a given galaxy we restricted the sample of detected point-like sources according to the following criteria: (1) in the case of large galaxies (e.g., NGC 1316 and NGC 4472), where part of the emission falls outside the S3 chip (in the adjacent front-illuminated chips), we ignored these outer sources, to avoid uncertainties introduced by the different instrumental responses. (2) We used only X-ray sources detected within the $D_{25}$ ellipse of each galaxy. While some sources found outside the $D_{25}$ ellipse could be associated with the galaxy under study, the contamination by background sources would be increased significantly by including these larger areas. Within the $D_{25}$ ellipse, instead, on average we expect only about 10% contamination by background sources, on the basis of the Log(N)-Log(S) relation of the ChaMP survey (Kim et al. 2004b). (3) We excluded sources found within 20″ of the galaxy center, to avoid contamination by nuclear sources, to minimize confusion effects in the crowded central regions, and to reduce the biases on the detection thresholds resulting from the more intense hot ISM emission of these inner regions. We note that this will not affect our results, because there is no systematic difference between radial distributions of X-ray bright and faint sources (Kim et al. 2004 in preparation),

With the above restrictions, we derive a total number of 985 sources from our sample of 14 galaxies, ranging from 30 to 140 sources per galaxy. The individual – uncorrected - cumulative XLFs are shown in Figure 1a. The location of $L_{X,Eddington} = 2 \times 10^{38}$ erg s$^{-1}$ is marked by vertical bars near the top and bottom of the figure. The XLFs appear to be following simple power-laws at the high $L_X$ end, and to flatten at the low $L_X$ side. As we will show below, this flattening is mostly the result of incompleteness effects.

3.1. Incompleteness correction

As shown by **KF03**, the apparent flattening of the XLFs at the low luminosities may stem from incompleteness effects. A full list of these effects includes: (1) the effect of the strong diffuse emission present in E and S0 galaxies (from hot ISM) on the detection probability for faint sources, (2) the Eddington bias, which causes a spurious apparent increase of sources detected near threshold, (3) confusion effects, and (4) the effect of the increasing PSF at larger radii on the source detection thresholds.

To assess the correction to be applied to each *observed* XLF, we have followed the 'backward' method outlined in the Appendix B of **KF03**, which corrects all the above effects at the same time. This method consists of running a series of simulations for each galaxy, adding one source at a time to the real observational data within the defined spatial region (r > 20″ and within the $D_{25}$ ellipse), and determining whether the inserted source could be detected by the same technique used for the *Chandra* data. We ran 20,000 simulations per galaxy. The 'added' source was selected with a random $L_X$ based on a single power-law XLF, with $\alpha = 1$ in a cumulative form,

$$N(>L_X) \sim L_X^{-\alpha}$$

and placed in a random location based on the optical light distribution assuming the $r^{1/4}$ law (de Vaucouleurs 1948; see also section 5.3). Note that the input XLF used for these simulations is not critical, because we are only using the ratio of the number of input to output (i.e., detected) sources to estimate the correction to the XLFs at each luminosity. The corrected XLFs are shown in Figure 1b. The XLF shapes are all remarkably similar apart from the factor ~20 spread in normalization at $L_X \sim 10^{38}$ erg s$^{-1}$. The apparent strong XLF breaks near $L_{X,Eddington}$ visible in Figure 1a mostly disappear after the corrections are applied.

3.2 The shapes of individual XLFs

Figure 2 (same as Figure 1b but rescaled by arbitrary factors for visibility) shows a variety of XLF shapes in our 14 galaxy sample. Is this variety due to intrinsic differences in the XLFs, or is it driven by small number statistics? Remember that 30-140 sources are detected in each galaxy. To address this question, we again resorted to simulations, generating 100 simulated XLFs selected from the same parent population. From the 20,000 simulations generated for NGC 1399 with a single power law with a slope $\alpha=1$ (see Section 3.1), we selected 100 sets of 100 detected sources randomly. The XLFs derived from these simulations are then corrected by the ratio of input and detected number of sources as discussed in Section 3.1. The corrected XLFs are plotted in Figure 3 (we plot only 10 XLFs for visibility), which can be directly compared with Figure 1b and Figure 2. As expected, the apparently XLF flattening at lower luminosities disappears in the corrected XLFs. In spite of the fact that they were all drawn from the same parent population, the simulated XLFs of Figure 3 have different shapes. This apparent variety appears consistent with that observed in the real XLFs, and simply reflects the limitation of small number statistics.

We have fitted the XLFs with a variety of methods to derive best-fit parameters, goodness of fit, and confidence intervals. These methods include: Maximum-likelihood (ML; Crawford et al. 1970), the Kolmogorov-Smirnov (K-S) test, minimum-$\chi^2$, c-statistic (cstat) and Cash (the last three are available in the CIAO Sherpa package). Cash and cstat are similar to the ML method. Of these methods, the minimum-$\chi^2$ method is best-calibrated (hence most extensively used, for example in X-ray spectral fitting) and can provide best-fit parameters, confidence intervals and an estimate of the goodness of the fit. However, it requires binned data. The other methods work with unbinned data, but only provide limited statistical information (e.g., best-fit parameters only for ML and goodness only for KS).

Since the errors in the cumulative XLF are not independent, we apply the statistical tests to the differential XLF given by

$$\frac{dN}{dLx} = K\, Lx^{-\beta}$$

For a single power-law, $\beta = \alpha + 1$ where $\alpha$ is the power-law index of the cumulative XLF (defined in section 3.1), while for a broken power-law, $\beta = \alpha + 1$ only hold above the break luminosity and $\beta < \alpha + 1$ below the break. We also note that confidence ranges determined by cstat and Cash statistics are not reliable because the corrected XLFs do not follow Poisson statistics.

The results of the fit are summarized in Table 2. The errors quoted in the table (and throughout this paper) are at the 90% confidence level, unless specified otherwise. The results of cstat and Cash are similar to those of ML and are not listed. With the KS test, we estimated the best-fit parameter via a grid-search method, as described in Fasano et al. (1993). Because the statistic (D) in the KS test is not fully calibrated and the best-fit parameter determined by the KS method may not be reliable, we only use the KS results in conjunction with other test results. For the minimum-$\chi^2$ fits, we binned the data to have at least 10 (uncorrected) sources in each bin.

The bias-corrected XLF of individual galaxies is well reproduced by a single, unbroken power-law. The differential best-fit slopes are, $\beta = 2.0 \pm 0.2$ for most galaxies. In a few galaxies with smaller numbers of sources (< 50), the slope is slightly flatter, $\beta = 1.7 \pm 0.4$, but still consistent with those of the main group. We note that none of the XLF fits can formally exclude a single power-law (the probability in column 4 and 8 in Table 2 is always larger than 20%). NGC 4697 is the only galaxy with an XLF, which may suggest (but not require) a more complex model than a single power-law. However, the statistics is still marginal (6% and 46% in the KS and minimum $\chi^2$ test, respectively).

For comparison, we have fitted the simulated XLFs in the same way as the real XLFs in Table 2. This test also allows us to estimate the size of systematic errors in our approach. Note that the simulated XLFs were all made with 100 sources (before bias-correction) taken from the same parent population. The best-fit slope of the 100 simulated XLFs ranges from $\beta=1.8$ to $\beta=2.2$ with a typical error of ~0.2, or ~8% (at 90% confidence). Similarly, the best-fit amplitude varies by ~12% (at 90% confidence). In Figure 4 we compare the best-fit slopes determined by the minimum-$\chi^2$ method (column 5 in Table 2) for the simulated XLFs (Figure 3) and real bias-corrected XLFs (Figure 1b). The spread of the best-fit slope of simulated XLFs is similar to that of the real bias-corrected XLF, indicating that the spread in the real XLFs is within the range of expected error. Our result demonstrates that the apparent variety of XLF shapes could be an artifact of poor count statistics. Moreover, models more complex than single power-laws are not required by the data.

However, we find a significant effect when we compare the amplitudes of the XLFs, which is not surprising since the same galaxy was used for generating the simulated XLF, while the real XLF reflects a range of galaxian population sizes. The amplitude of the XLF (column 6), which is a measure of the total number of X-ray sources, varies considerably by a factor of ~20 from one galaxy to another (or by a factor of ~3 after normalized by $L_K$; see section 4.0). For comparison, the amplitude in the simulation test varies only within ~12%. We will discuss the implications of this result for the scatter of $L_X$(LMXB) and its ratio to the optical luminosity in Section 4.0.

3.3. Combined XLF

As shown in Section 3.2, all the bias-corrected XLF slopes are consistent within the errors. We therefore combined them, to increase the statistical significance of our data. Because the range of $L_X$ covered by the XLF is different from one galaxy to another, we used the combined XLF only in the luminosity range $L_X > 6 \times 10^{37}$ erg s$^{-1}$ for which all the galaxies give detections. This combined XLF includes 874 sources and is plotted in Figure 5. In fitting the combined XLF, we take into account various errors. First, we determine the statistical error by Poisson statistics. Second, we add the systematic error estimated in section 3.2, which is 12% at 90%, or 7% at 1$\sigma$. Third, we also add the distance error. While group distance errors for Virgo and Fornax galaxies are small (~2%), individual distance errors for our 14 galaxies are ~8% (Tonry et al. 2001). We have compared the two cases using group distances and individual galaxy distances, but the results are consistent with each other. The only noticeable difference (although still consistent within the error) is that the break luminosity, when a broken power-law model is used, is slightly higher (by ~10%) in the latter case with a slightly larger error. To properly consider these errors in determining the XLF parameters and their errors, we have used a Monte-Carlo technique by applying the statistical and systematic errors to the number of sources (i.e., vertical variation of XLF) and the distance error to $L_X$ (i.e., horizontal variation of XLF). Note that applying the systematic and distance errors in addition the statistical error does not significantly change the best-fit parameters, but makes their acceptable ranges (or errors) larger.

A fit with a single power-law of this XLF (Figure 5b) yielded the best-fit single power-law slope, $\beta = 2.1 \pm 0.1$, consistent with the individual XLF fits, but with a smaller error. Formally, a single power-law is marginally acceptable with typical $\chi^2_{reduced} = 2.0$ for 13 degrees of freedom (~1% probability of exceeding the $\chi^2$ statistic).

Fitting the XLF with a broken power-law model results in a better fit, with $\chi^2_{reduced}$ close to 1 for 11 degrees of freedom (Figure 5b). The F-test indicates that the probability of exceeding F is ~1%. The best-fit break luminosity is $5.0 \pm 1.6 \times 10^{38}$ erg s$^{-1}$. The slope of the low-luminosity power-law is $\beta = 1.8 \pm 0.2$, while that of the high luminosity power law is $\beta = 2.8 \pm 0.6$. Note that the cumulative XLF slope below the break is similar to that obtained from the single power-law model ($\alpha \sim 1.1$), because $\beta < \alpha + 1$ below the break (see Figure 7 and section 5.2). The effect of this break is not a flattening of the low-luminosity XLF (compared to a single power-law fit), but a steepening of the XLF at higher luminosities: note that the high luminosity slope is more uncertain, given the small number of very bright sources. This point is particularly important for the determination of the contribution from LMXBs to the integrated X-ray luminosity of the galaxies and will be discussed in Section 4 and 5.

The break suggested by the fit of the combined XLF occurs at a higher luminosity (by a factor of ~2.5) than $\boldsymbol{L_{X,Eddington}} = 2 \times 10^{38}$ erg s$^{-1}$ (with a ~3$\sigma$ significance; see section 5.1). We exclude that the break luminosity could be the result of the wrong spectral assumptions producing apparently larger X-ray luminosities for these bright sources (see Finoguenov and Jones 2002), if they have a softer X-ray spectrum than assumed in the count rate to flux conversion. The spectral properties for sources with different $L_X$ are remarkably similar in NGC 1316 (**KF03)** and also in a large sample of early type galaxies (Irwin et al. 2003); all these spectra tend to be rather hard.

We note that cosmic background X-ray sources, which contribute to ~10% of the detected sources (Table 1, column 17) in our LMXB flux range (and within the $D_{25}$ ellipse), exhibit a similar break in the Log(N)-Log(S) relation (e.g., Kim et al. 2004b; Hasinger et al. 1998). At the distance of the Virgo cluster (17 Mpc), the break luminosity corresponds to ~$5 \times 10^{38}$ erg s$^{-1}$, similar to the location of the XLF break. To assess quantitatively the effect of the cosmic background X-ray sources, we added an extra broken power-law component, based on the ChaMP Log(N)-Log(S) by Kim et al (2004b), to the above fitting. The results are statistically consistent within the error. Irwin et al. (2003) have recently suggested that very bright sources ($L_X > 2 \times 10^{39}$ erg s$^{-1}$) found in early type galaxies may be background AGNs. We have repeated our fitting with only sources of $L_X < 2 \times 10^{39}$ erg s$^{-1}$ and again reached consistent results within the error.

**4.0 Total Integrated X-ray Luminosity of the LMXB Population**

Using the individual XLFs, we can estimate the integrated X-ray luminosity of both detected and undetected sources in our sample of galaxies. Given the old stellar populations of these systems, point-like X-ray sources are likely to be entirely representative of LMXBs. Our results are summarized in Table 3. Because the XLF is steep ($\beta \sim 2.1$), the total X-ray luminosity of LMXBs depends on the undetected lower $L_X$ break of the XLF, which in our sample is below a few x $10^{37}$ erg s$^{-1}$. Following LMXBs detected in the Milky Way (Grimm et al. 2002) and in the bulge of M31 (Kong et al. 2002), we assume the lower limit to the XLF at $10^{37}$ erg s$^{-1}$, and list the result in column (2) of Table 3. If we conservatively take a larger limit of $4 \times 10^{37}$ erg s$^{-1}$, which is the observational threshold of our sample, $L_X$ decreases by 20-25%. In column (3) we list a correction factor that takes into account the omission of sources from the XLFs, either because they did not fall in the S3 chips, or because of the exclusion of the central 20" of the galaxies from the XLF (see section 3). We assume that LMXBs are spatially distributed following the optical light ($r^{1/4}$ law). The corrected total LMXB luminosities are listed in column (4). These luminosities range from a few x $10^{39}$ to several x $10^{40}$ erg s$^{-1}$.

Table 3 lists the ratio $L_X$(LMXB)/$L_B$, in unit of $10^{40}$ erg s$^{-1}$ / $10^{10}$ $L_{B\odot}$, where $L_B$ represents the integrated stellar luminosity in the optical band. $L_B$ was derived from $B_T^o$, the total face-on magnitude corrected for galactic extinction (column 8 in Table 1, taken from **RC3**), adopting $M_{B\odot}$=5.47 mag. The X-ray to optical (B) luminosity ratios range from 0.3 to 1.7 (with the lowest in NGC 4382 and the highest in NGC 1399). The mean and standard deviation are:

$L_X$(LMXB)/$L_B$ = 0.9 ± 0.5 x $10^{30}$ erg s$^{-1}$ / $L_{B\odot}$   with $L_X$(min) = $10^{37}$ erg s$^{-1}$.

Our estimates are consistent with the previous measurements by Kim et al. (1992) and O'Sullivan et al. (2001), after correcting for the different energy bands used. However, the large scatter is statistically significant, if the XLFs of individual E and S0 galaxies do not differ significantly below the detection threshold.

Also in Table 3 we list the ratio $L_X$(LMXB)/$L_K$. We have introduced the K-band luminosity ($L_K$) as an additional measure of the integrated stellar emission, because the near-IR luminosity may be more appropriate for the old stellar population of early type galaxies and also is less affected by extinction than the B-band luminosity. $L_K$ was derived from $K_{20}$ measured within the 20 mag arcsec$^{-2}$ isophote (column 9 in Table 1, taken from 2MASS; Jarrett et al. 2003), and using $M_{K\odot}$ = 3.33 mag. With $L_K$, we obtain:

$L_X$(LMXB)/$L_K$ = 0.20 ± 0.08 x $10^{30}$ erg s$^{-1}$ / $L_{K\odot}$   with $L_X$ (min) = $10^{37}$ erg s$^{-1}$.

Figure 6 show the scatter diagram between $L_X$(LMXB) against $L_K$ for our 14 galaxies. Two extreme galaxies, NGC 1399 (with the highest $L_X$(LMXB)/$L_K$) and NGC 3379 (with the lowest $L_K$), are marked by a large circle and a large square, respectively. This figure shows that there is considerable spread in X-ray-near-IR ratios. This scatter exceeds that expected from the uncertainties in the measurement of $L_X$(LMXB). The scatter of $L_X$(LMXB)/$L_K$ (Figure 6) is ± 40% at 1σ rms. If we use $L_B$, the scatter is slightly larger, ±60% at 1σ rms, in a good agreement with a factor of 4 scatter (in a full width) reported by White et al. (2002). We will further discuss the $L_X$(LMXB)/$L_K$ scatter in Section 5.3.

Nevertheless, significant correlations exists between $L_X$(LMXB) and both $L_B$, and $L_K$ (see Figure 6). The Spearman Rank correlation coefficients are 0.71 and 0.82, corresponding to chance probabilities of 1% and 0.3% for the B and K correlations, respectively, confirming the tighter correlation with $L_K$. Also plotted in Figure 6 is the bulge of the Milky Way (star) and M31 (asterisk). For the Milky Way, the optical luminosity of the bulge was taken from Cox (1999) and the X-ray luminosity of LMXBs from Grimm et al. (2003), while for M31, the optical luminosity is from Kent (1989) and the X-ray luminosity from Trinchieri and Fabbiano (1991). $L_X$ was rescaled for the energy range of 0.3 – 8 keV. The near-IR luminosity ($L_K$) was estimated using the average B-K color (or $L_K/L_B$ = 4.5) in our sample of early type galaxies. Although these quantities are somewhat uncertain, [e.g., Widrow et al. (2003) suggested a lower (by a factor of 2) M31 bulge mass than measured by Kent (1989)], the bulges of the Milky Way and M31 follow the general trend of early type galaxies and roughly indicate the upper and lower bounds of the $L_X/L_K$ scatter (see section 5.2 for more comparisons).

**5. Discussion**

5.1 The XLF of early-type galaxies

The study of the X-ray source populations of early type galaxies detected with *Chandra* has generated a considerable amount of excitement and some controversy. Excitement because we finally have irrefutable proof of the existence of these sources (e.g. Sarazin et al. 2000), which had been predicted on the basis of indirect evidence since the first observations of early-type galaxies with the *Einstein Observatory* (Trinchieri & Fabbiano 1985; see Fabbiano 1989). Controversy because of the derivation and interpretation of the XLFs of these X-ray source populations: the early suggestion (e.g., Sarazin et al. 2000; Blanton et al. 2001) of a universal break in the XLFs at the Eddington luminosity of neutron stars ($L_{X,Eddington}$ ~ 2 x $10^{38}$ erg s$^{-1}$), was later related to the lack of careful incompleteness corrections (**KF03**, for NGC 1316; see also Sivakoff et al. 2003, for NGC 4365 and NGC 4382). The lack of a signature at the Eddington luminosity is intriguing because this effect is expected if the emission of X-ray binaries is Eddington-limited, as demonstrated by synthetic XLFs based on binary evolution models (Kalogera et al. 2003).

The purpose of the work reported in this paper was to address the issue of the shape of the XLF, by performing a rigorous analysis of a representative sample of E and S0 galaxies, leading to a set of uniformly derived completeness-corrected XLFs. As reported in Section 3, to a first approximation all these XLFs are well fitted with single power-law models, with best-fit (differential) slopes ranging from 1.7 to 2.2. Comparing these results with a set of synthetic XLFs, we conclude that this apparent variety may be the result of small number statistics (typically 100 sources or less are involved in each XLF). We do not find any strong statistical evidence of breaks at $L_{X,Eddington}$ in these XLFs, and we therefore conclude that incompleteness played a role in the early reports of this feature (e.g., Sarazin et al. 2000; Blanton et al. 2001).

Nonetheless, we find evidence of a break when we create a composite XLF for the entire sample, under the working assumption that there is a 'universal' XLF shape of the X-ray source population of old stellar systems. This composite XLF may suggest a departure from a simple power-law distribution, consistent with having a break at a luminosity of 5 x $10^{38}$ erg s$^{-1}$. If this break is real, its luminosity may be inconsistent with $L_{X,Eddington}$ of 1.4 M$_\odot$ neutron stars (at 3$\sigma$ confidence), but would be consistent with the 3 M$_\odot$ upper limit on the neutron star mass, suggested by Kalogera & Baym (1996). This break could also correspond to the Eddington luminosity of a ~3.5M$_\odot$ accreting black hole. Another possibility for the 'super-Eddington' luminosity is He-enriched accretion which could effectively double $L_{X,Eddington}$ because of the smaller cross-section per unit mass (see e.g., Shakura and Sunyaev 1973; Grimm et al. 2003). Podsiadlowski et al. (2002) suggested that a large fraction of LMXBs might have started as intermediate mass (neutron star) X-ray binaries (IMXBs) with an initial mass of a companion star ~ a few M$_\odot$. Because these systems would be in a later evolutionary stage (than a typical LMXB with < 1 M$_\odot$), they could be hydrogen-deficient and helium-enriched, hence the Eddington accretion rate may be considerably enhanced. However, these systems are relatively short-lived (Podsiadlowski et al. 2002) and while they may be present in young E/S0 galaxies such as Fornax A and Cen A, are not likely to be found in the old elliptical galaxies of our sample (Trager et al. 2000).

If the change in XLF slope is real, and does not mask a step in the XLF, that is not detectable because of the relative low number of high luminosity sources, our result would imply that there are two different source populations: neutron star and black-hole binaries, with different XLFs (see Sarazin 2000). A step in the XLF would be more indicative of a beaming effect, that may enhance the luminosity of some high accretion rate binaries (King 2002). In the latter case, we may expect that the fraction of sources with `enhanced' apparent luminosity would have the same luminosity function slope of their parent population, the unbeamed LMXBs. If the above consideration continues to hold for future observations, the high luminosity portion of the XLF could reflect the mass function of back holes in elliptical galaxies. We note that this high luminosity population does not resemble that of the ultra-luminous sources detected in star-forming galaxies, where no break in the XLF is present, and where the power-law slope is also much flatter than in the older stellar system we are studying here (see Fabbiano & White 2003 and refs. therein). Analysis of a larger sample is needed to firmly establish the behavior of the high luminosity portion of the XLF of LMXBs.

## 5.2 Comparison with the LMXB XLFs of M31 and the Galaxy

Figure 7 compares our composite cumulative XLF with the XLFs of the LMXBs found in the bulge of M31 (Kong et al. 2002), those matched with the globular clusters of M31 (Kong et al. 2003) and with that of the LMXBs in the Milky Way (Grimm et al. 2002). The similarity of the shapes of these three luminosity functions, in the luminosity range covered by our XLF, is remarkable, and suggests that we are indeed looking at similar X-ray binary populations. The cumulative power-law slopes determined by Kong et al (2002, 2003) for the bulge of M31 and for the globular clusters of M31 ($\beta = 2.15$ at $L_X > 0.8 \times 10^{37}$ ergs s$^{-1}$ and $\beta = 2.2$ at $L_X > 2 \times 10^{37}$ ergs s$^{-1}$), compares well with our cumulative XLF, $\beta = 2.1 \pm 0.1$ for a single power-law, or $\beta = 1.8 \pm 0.2$ at luminosities below the break ($5 \times 10^{38}$ erg s$^{-1}$) for a broken power-law (note that even in the broken power-law fit, the cumulative XLF slope below the break is close to $\alpha = 1.1$). The LMXBs in the Milky Way (Grimm et al. 2002) show the same trend. A significant difference in both the M31 and Milky Way LMXB XLFs is the absence of the luminous sources ($L_X > 2 \times 10^{38}$ ergs s$^{-1}$), that we detect in E and S0 galaxies. However, this could be just a result of population statistics, given the smaller population sizes of the bulge of M31 and the Milky Way, when compared with our sample E and S0 galaxies. For example, for a given $L_B$ of the Milky Way bulge, we expect only a few LMXBs with $L_X > 10^{38}$ erg s$^{-1}$, consistent with observations (e.g., Grimm et al. 2002)

In both the M31 and the Milky Way LMXB XLFs, there is a low luminosity break near $L_X \sim 10^{37}$ erg s$^{-1}$, and the slope flattens considerably at lower luminosities. This luminosity range is well below the sensitivity of the *Chandra* observations used in this study. Only in NGC 3379, the nearest galaxy in our sample, can we detect sources with luminosities as low as $L_X = 2 \times 10^{37}$ erg s$^{-1}$ (see the bottom XLF in Figures 1 and 2), and in this case the low luminosity break is not seen. Given the similarity between all these XLFs, it is reasonable to assume that there may be a low luminosity break in the XLFs of E and S0 at a similar luminosity as seen in M31 and the Milky Way. Based on the LMXBs found in the Milky Way (Grimm et al. 2002), the integrated X-ray luminosity of LMXBs with $L_X < 10^{37}$ erg s$^{-1}$ is only ~8% of the total LMXB emission.

## 5.3 The $L_X$(LMXB)/$L_{opt}$ relation and the evolution of LMXBs

With the detection of X-ray source populations in early-type galaxies there has been renewed interest in probing the formation of these systems, and in particular in exploring a possible evolutionary link to Globular Clusters (GCs), originally suggested by Grindlay (1984) as the main formation mechanism for Galactic LMXBs. A significant fraction of the sources detected with *Chandra* in early-type galaxies are associated with GCs (e.g. Angelini et al 2001; Kundu, Maccarone & Zepf 2002; see compilation in Fabbiano & White 2003). This association has led to the suggestion that GCs may be the birthplace of the entire LMXB population, from where they may be expelled if they receive strong enough formation `kicks', or may be left behind upon tidal disruption of the parent cluster (see Sarazin, Irwin & Bregman 2000; White, Sarazin & Kulkarni 2002).

Following White et al. (2002), who suggested a correlation of the $L_X(LMXB)/L_{opt}$ ratio with the GC specific frequency, we explored the behavior of the $L_X(LMXB)/L_{opt}$ ratio of our sample, to see if we could find any trends. Our advantage is that our study of the XLF should result in the most rigorous determination of the integrated LMXB luminosities. Furthermore, because we find that the shape of LMXB XLFs is uniform, the correlation, if confirmed, directly indicates that the number of LMXBs is proportional to the number of globular clusters in individual E/S0 galaxies. This would in turn provides strong evidence that LMXBs are mainly, if not all, formed in globular clusters.

While the global stellar content (measured by the optical or near IR light) is correlated with the integrated luminosity of LMXBs (Figure 6), the scatter seems to require a secondary parameter. Using the $L_X(LMXB)/L_K$ ratio, which has the least amount of scatter (see Section 4), we explore dependencies of the scatter from a second parameter in Figure 8, where we plot this ratio versus $L_K$, and versus the GC specific frequency, $S_{GC}$, defined by a number of globular clusters per unit galaxy luminosity:

$$S_{GC} = N_{GC} \times 10^{0.4(M_V + 15)}.$$

$S_{GC}$ was estimated from the globular cluster data listed in Table 1. NGC 1407, NGC 4382 and IC 1459 do not have $S_{GC}$ data and are marked by triangles at the left side of Figure 8. For comparison, we plot the Milky Way (and M31), by rescaling $S_{GC} = 0.5$ (and 0.7) of the whole Galaxy (Harris 1991) for the bulge luminosity (Cox 1999; Kent 1989). However, we note that this is subject to a large uncertainty because this quantity may not be simply scalable (e.g., by local/global variations).

Figure 8a shows that there is no dependency of $L_X(LMXB)/L_K$ on the integrated stellar luminosity ($L_K$). Instead (Figure 8b) there is a suggestion of a correlation with $S_{GC}$, which would follow the suggestion of White et al (2002). The Spearman Rank correlation test gives a chance probability of 2%. However, if we exclude NGC 1399, which is the galaxy with the largest $S_{GC}$ (marked by a large circle at the top-right corner in Figure 9b), the chance probability increases to 5%. We note that another known elliptical galaxy with a very high $S_{GC}$, NGC 4486, which hosts twice more globular clusters than NGC 1399 (Kissler-Patig 1997), does not have as many LMXBs as in NGC 1399 (Kim et al. 2004 in preparation), indicating NGC 1399 may not follow the general trend of early type galaxies. We also note that the measured $S_{GC}$ vary considerably, for example $S_{GC}$ of NGC 4374 differs by a factor of 2.4 between Kissler-Patig (1997) and Gomez and Richtler (2003). In conclusion, although there is a weak suggestion of a connection between $S_{GC}$ and $L_X(LMXB)/L_B$, this is by no means an established result. The LMXB – globular cluster connection is still an open issue that needs to be confirmed with a larger sample of galaxies with well-determined $S_{GC}$.

Based on the GC-LMXB connection, one might expect the radial distribution of LMXBs is similar to that of GCs, hence flatter than that of the optical halo light. However, using ground and HST observations of 6 giant elliptical Galaxies, Kim, E. et al. (2004; in preparation) found that LXMBs closely follow the optical halo light than more extended GCs, suggesting a rather complex connection, if any, depending on various factors operating in the LMXB formation in GCs and its subsequent evolution. The close agreement between the radial distributions of the optical light and LMXBs is also seen in NGC 1316 (**KF03**). This justifies our usage of $r^{1/4}$ law for the LMXB distribution (section 3.1 and 4). As already suggested by Kundu et al. (2002) and Sarazin et al. (2003), Kim, E. et al. (2004) confirmed a significantly higher probability (by a factor of 4) to find LMXBs in red (metal-rich) GCs than in blue (metal-poor) GCs, indicating that the metal abundance may play a key role in forming LMXBs (e.g., by flatter IMF or by irradiation induced stellar winds as suggested by Maccarone, et al. 2004). The observed radial behavior of LMXBs may be partly understood, because red GCs are often more centrally concentrated than blue GCs (Lee et al. 1998; also Kim E. et al. 2004). However, there may be other factors, such as variable capture rates as a function of cluster density, which could make the LMXB distribution look steeper.

6. Conclusions

We have derived bias-corrected XLFs in the range $L_X = 2 \times 10^{37} - 2 \times 10^{39}$ erg s$^{-1}$ for a uniformly selected sample of 14 E and S0 galaxies observed with *Chandra* ACIS-S3, following the approach of **KF03**. The entire sample yields 985 point-like X-ray sources, with 30-140 sources detected per galaxy. Our analysis shows that:

1. After correcting for incompleteness, the individual XLFs are statistically consistent with a single power-law of a (differential) slope $\beta = 1.8 - 2.2$. A break at or near $L_{X,Eddington}$, as reported in the literature for some of these galaxies (see review in Fabbiano & White 2003) is not required in any case. We also demonstrate, by comparison with simulated data generated from the same parent power-law XLF, that the apparent differences in the shape of each individual XLF are consistent within statistics.

2. Given that all the XLFs have statistically consistent shapes, we have combined the entire data set, to generate a representative XLF of X-ray sources in elliptical galaxies, with increased sensitivity. Although the combined XLF is marginally consistent with a single power-law, a broken power-law gives an improved fit. The best-fit (differential) slope is $\beta = 1.8 \pm 0.2$ in the low-luminosity range $L_X =$ a few $\times 10^{37} - 5 \times 10^{38}$ ergs s$^{-1}$. At higher luminosities, the slope is steeper $\beta = 2.8 \pm 0.6$. The break luminosity is $5 \pm 1.6 \times 10^{38}$ ergs s$^{-1}$. We note that the low-luminosity slope (in a cumulative form) is consistent with that obtained from the single power-law model ($\alpha = 1.1 \pm 0.1$; $\beta < \alpha + 1$ below the break). We find that the break luminosity is not significantly altered by either spectral or distance uncertainties in our sample, or by contamination by background AGNs.

3. If the XLF is indeed broken, the break luminosity may be higher than $L_{X,Eddington}$ of a 1.4 M$_\odot$ neutron star (at ~3 $\sigma$ confidence), but consistent with the Eddington luminosity of the largest possible neutron star mass proposed (3 M$_\odot$, Kalogera & Baym 1996). This break luminosity is in the range of $L_{X,Eddington}$ of stellar mass black holes. Alternatively, a *He*-enriched LMXB could also be consistent with a larger $L_{X,Eddington}$. If the change in slope is real, and does not mask a step in the XLF, that is not detectable because of the relative low number of high luminosity sources, our result would imply a different population of high luminosity sources, instead of a beaming effect. This high luminosity portion of the XLF could reflect the mass function of black holes in these galaxies. We note that this high luminosity population does not resemble that of the ultra luminous sources detected in star-forming galaxies, where no break in the XLF is present. Also, the power-law slope is much flatter in star-forming galaxies than in the older stellar system we are studying here (see Fabbiano & White 2003 and refs. therein).

4. Within statistics, our low-luminosity composite XLF is fully consistent with the XLFs of LMXBs of the Milky Way (Grimm et al 2002) and of both the bulge and globular clusters of M31 (Kong et al 2002, 2003). The proximity of the Milky Way and M31 sources allows a measurement of their XLFs down to significantly lower luminosities, demonstrating that the single power-law (with $\beta \approx 2.2$) continues down to $L_X \approx 10^{37}$ erg s$^{-1}$.

5. In contrast to the uniform XLF slope, the normalization of the XLFs varies widely from one galaxy to another, reflecting the varying content of LMXBs. The total X-ray luminosity of LMXBs is correlated both with optical and (better) near-IR luminosities, but in both cases the scatter exceeds that expected from measurement errors. The regression lines are:

$L_X(LMXB)/L_B = 0.9 \pm 0.5 \ \ \text{x} \ 10^{30}$ erg s$^{-1}$ / $L_{B\odot}$ and

$L_X(LMXB)/L_K = 0.2 \pm 0.08 \ \text{x} \ 10^{30}$ erg s$^{-1}$ / $L_{K\odot}$

6. Following White et al (2002), we find that the scatter in $L_X(LMXB)/L_K$ is marginally correlated with the specific frequency of globular clusters ($S_{GC}$), suggesting an important role of globular clusters in LMXB evolution. This conclusion, however, needs to be confirmed with a larger sample of galaxies with accurately measured $S_{GC}$

After submitting this paper, we became aware that Gilfanov (astro-ph/0309454) reached consistent conclusions (on the XLF shape and the break luminosity) with different samples and by independent approaches.

AKNOWLEDGEMENTS

This work was partly supported by NASA grant NAS8-39073 (CXC). We thank H.-J. Grimm for providing the Milky Way data, V. Kashyap for discussing statistics issues and M. Elvis for carefully reading the draft.

Figure Captions

Figure 1. (a) Uncorrected XLFs of LMXBs for 14 early type galaxies listed in Table 1. (b) Bias-corrected XLFs of LMXBs. The location of the Eddington luminosity ($2 \times 10^{38}$ erg s$^{-1}$) of a 1.4 M$\odot$ neutron star is marked by vertical bars.

Figure 2. Same as Figure 1b, but rescaled by arbitrary factors (second numbers in parentheses below) for visibility. From top to bottom in (b) are NGC1399 (red, 50.12), NGC4649 (black, 39.81), NGC4472 (blue, 29.51), NGC4636 (green, 50.12), NGC4365 (black, 17.78), NGC1407 (black, 1.26), NGC1316 (black, 1.26), NGC720 (blue, 0.50), NGC4697 (black, 0.63), IC1459 (red, 0.19), NGC4382 (green, 0.32), NGC4374 (blue, 0.12), NGC4621 (green, 0.11), and NGC3379 (red, 0.18). The location of the Eddington luminosity ($2 \times 10^{38}$ erg s$^{-1}$) of a 1.4 M$\odot$ neutron star is marked by a vertical line.

Figure 3. Ten simulated XLFs with 100 sources drawn from the same parent population of an XLF (differential) slope of 2 after correction. They are directly compared with Figure 1b and 2. They are arbitrarily rescaled for visibility.

Figure 4. Distribution of the best-fit XLF slopes for (a) simulations and (b) observations. The simulated XLFs (a) are made with 100 LMXBs, while those observed (b) with 30-140 LMXBs.

Figure 5. Fit of the combined (differential) XLF to (a) a single power-law, and (b) a broken power-law.

Figure 6. The X-ray luminosity of LMXBs, $L_X$(LMXB), is plotted against (a) the optical luminosity $L_B$ and (b) the near-infrared luminosity $L_K$. The dashed line indicates the linear relation (slope = 1.0). Two extreme galaxies, NGC 1399 (with the highest $L_X(LMXB)/L_B$) and NGC 3379 (with the lowest $L_B$), are marked by a large circle and a large square, respectively. Three galaxies with no globular cluster data are marked with triangles (to be consistent with Figure 8). A typical error bar is shown in the top left corner. Also plotted is the Milky Way (marked by a star) in the bottom left corner (see text).

Figure 7. Cumulative XLF of the 14 E and S0 galaxies. The long-dashed and solid lines are the best-fit single power-law and broken power-law. The vertical bar at the top indicates the break $L_X$ in the broken power-law-fit. Also plotted at the lower left corner are XLFs determined with LMXBs in the Milky Way (squares with error bar) and M31 (dotted and dashed lines are for the Bulge and GC sources).

Figure 8. $L_X(LMXB)/L_K$ is plotted against (a) $L_K$ and (b) the globular cluster specific frequency ($S_{GC}$). Two extreme galaxies, NGC 1399 (with the highest $L_X(LMXB)/L_B$) and NGC 3379 (with the lowest $L_B$), are marked by a large circle and a large square, respectively. Three galaxies with no $S_{GC}$ data are marked with triangles. Also plotted is the Milky Way (marked by a star) (see text).

Table 1   Sample Galaxies

| name | obsid | obs date | N(H) | ECF | D (Mpc) | type (T) | D_25 (') | BTo (mag) | K_20 (mag) | S(GC) | exp (sec) after/before screening | | # source S3 | D25 | r>20" | Bkgd | ref |
|---|---|---|---|---|---|---|---|---|---|---|---|---|---|---|---|---|---|
| (1) | (2) | (3) | (4) | (5) | (6) | (7) | (8) | (9) | (10) | (11) | (12) | (13) | (14) | (15) | (16) | (17) | (18) |
| I1459 | 2196 | Aug 12 2001 | 1.17 | 72.958 | 29.0 | −5 | 5.25 | 10.83 | 6.929 | − | 53055 | 60166 | 74 | 59 | 44 | 5 | 1 |
| N0720 | 492 | Oct 12 2000 | 1.57 | 69.810 | 28.0 | −5 | 4.68 | 11.13 | 7.393 | 1.0 | 32802 | 40124 | 69 | 36 | 30 | 2 | 2 |
| N1316 | 2022 | Apr 17 2001 | 2.13 | 72.202 | 19.9F | −2 | 12.02 | 9.40 | 5.688 | 2.5 | 24683 | 30233 | 83 | 77 | 66 | 10 | 3 |
| N1399 | 319 | Jan 18 2000 | 1.50 | 63.209 | 19.9F | −5 | 6.92 | 10.44 | 6.440 | 8.6 | 55645 | 56659 | 181 | 142 | 139 | 6 | 4 |
| N1407 | 791 | Aug 16 2000 | 5.43 | 77.391 | 29.0 | −5 | 4.57 | 10.71 | 6.855 | − | 42352 | 49196 | 146 | 87 | 82 | 5 | |
| N3379 | 1587 | Feb 13 2001 | 2.78 | 74.407 | 11.0 | −2 | 5.37 | 10.18 | 6.362 | 1.1 | 28851 | 31923 | 69 | 44 | 27 | 5 | |
| N4365 | 2015 | Jun  2 2001 | 1.61 | 73.200 | 17.0V | −5 | 6.92 | 10.49 | 6.800 | 6.3 | 38748 | 40947 | 135 | 112 | 101 | 7 | 5 |
| N4374 | 803 | May 19 2000 | 2.78 | 69.746 | 17.0V | −5 | 6.46 | 10.01 | 6.347 | 2.4 | 27587 | 28841 | 86 | 65 | 54 | 7 | 6 |
| N4382 | 2016 | May 29 2001 | 2.50 | 75.163 | 17.0V | −1 | 7.08 | 9.99 | 6.260 | − | 37477 | 40259 | 71 | 50 | 41 | 8 | 5 |
| N4472 | 321 | Jun 12 2000 | 1.62 | 67.578 | 17.0V | −5 | 10.23 | 9.33 | 5.506 | 3.9 | 33568 | 40096 | 136 | 129 | 123 | 10 | 7 |
| N4621 | 2068 | Aug  1 2001 | 2.17 | 75.073 | 17.0V | −5 | 5.37 | 10.53 | 6.866 | 5.5 | 23354 | 25155 | 72 | 42 | 30 | 3 | |
| N4636 | 323 | Jan 26 2000 | 1.82 | 64.208 | 17.0V | −5 | 6.03 | 10.43 | 6.628 | 6.5 | 42408 | 53049 | 101 | 65 | 61 | 4 | 8 |
| N4649 | 785 | Apr 20 2000 | 2.13 | 67.607 | 17.0V | −5 | 7.41 | 9.70 | 5.825 | 6.0 | 21536 | 37350 | 141 | 119 | 116 | 6 | 9 |
| N4697 | 784 | Jan 15 2000 | 2.14 | 64.699 | 12.0 | −5 | 7.24 | 10.07 | 6.502 | 4.4 | 36663 | 39763 | 91 | 73 | 70 | 7 | 10 |

col 1 : galaxy name
col 2 : Chandra observation id
col 3 : Chandra observation date
col 4 : galactic N(H) in unit of $10^{20}$ cm$^{-2}$ (taken from Stark et al. 1992).
col 5 : energy conversion faction in unit of $10^{-13}$ erg sec$^{-1}$ cm$^{-2}$ per 1 cnt/sec
col 6 : distance in Mpc (Torny et al. 2001). Note that for galaxies in the Fornax (F) and Virgo (V) clusters, we use the group distance.
col 7 : morphological type (taken from RC3)
col 8 : diameter at the 25-th mag isophote in arcmin (taken from RC3)
col 9 : total face-on B magnitude corrected for galactic extinction (taken from RC3)
col 10: 2MASS K mag within the 20 mag arcsec$^{-2}$ (Jarret et al. 2003)
col 11: specific frequency of globular clusters taken (and corrected for our distances) from Goudfrooij et al. (2001, MNRAS, 328, 238)
       for NGC 1316, Rhode (2003, astro-ph/0310277) for NGC 3379 and NGC 4472, and Gomez and Richtler (2003;

astro-ph/0311188) for NGC 4374 and from Kissler-Patig (1997) for the rest galaxies. There is no available data for NGC 1407, NGC 4382 and IC 1459.

col 12: effective exposure time before excluding background flares
col 13: effective exposure time after excluding background flares
col 14: number of sources in the S3 chip
col 15: number of sources within the D25 ellipse in the S3 chip
col 16: number of sources at r>20" within the D25 ellipse in the S3 chip
col 17: number of expected background sources in the same region as in column 16 (based on logN-logS in Kim et al. 2004b)
col 18: References to previous works on the individual Chandra obs.

   1  I1459 Fabbiano et al. 2003, ApJ, 588, 175
   2  N0720 Jeltema, T. E., Canizares, C. R., Buote, D. A., & Garmire, G. P. 2003, ApJ, 585, 756
   3  N1316 KF
   4  N1399 Angelini eta l. 2001, ApJ, 557, L35
      N1407 none
      N3379 none
   5  N4365 Sivakoff et al. 2003
   6  N4374 Finoguenov, A., & Jones, C. 2002, ApJ, 574, 754
   5  N4382 Sivakoff et al. 2003
   7  N4472 Kundu, A., Maccarone, T. J., & Zepf, S. E. 2002, ApJ, 574, L4
      N4621 none
   8  N4636 Jones, et al. 2002, ApJ, 567, L115 (no discussion about LMXB)
   9  N4649 Randall, S. W., Sarazin, C. L., & Irwin, J. A. 2003, astro-ph/0309809
  10  N4697 Sarazin, C. L., Irwin, J, A., & Bregman, J. N. 2000, ApJ, 544, L101

Table 2 Fitting observed XLFs

| name | ML | KS | | chi-square method | | | |
|---|---|---|---|---|---|---|---|
| | beta | beta | prob(%) | beta | ampl | chi2_red (chi2 /dof) | prob(%) |
| (1) | (2) | (3) | (4) | (5) | (6) | (7) | (8) |
| I1459 | 1.66 | 1.80 | 39.24  | 1.70 (−0.30 +0.31) | 39.14 (−16.26 +19.07) | 0.39 ( 1.16 / 3) | 76.21 |
| N0720 | 1.96 | 1.97 | 100.00 | 1.62 (−0.83 +0.67) | 33.01 (−27.05 +57.79) | 0.42 ( 0.42 / 1) | 51.71 |
| N1316 | 2.16 | 1.97 | 96.52  | 2.13 (−0.23 +0.26) | 61.93 (−17.03 +17.52) | 0.25 ( 1.23 / 5) | 94.23 |
| N1399 | 2.02 | 1.95 | 84.66  | 2.00 (−0.16 +0.17) | 111.14 (−21.98 +22.48) | 0.97 (11.64 /12) | 47.48 |
| N1407 | 2.07 | 1.98 | 97.47  | 2.11 (−0.28 +0.29) | 132.01 (−49.96 +61.87) | 0.79 ( 5.54 / 7) | 59.41 |
| N3379 | 1.62 | 1.80 | 44.84  | 1.73 (−0.35 +0.40) | 6.73 ( −2.49  +2.34) | 1.07 ( 1.07 / 1) | 30.11 |
| N4365 | 1.94 | 1.94 | 82.95  | 1.94 (−0.18 +0.20) | 38.57 ( −6.73 + 6.72) | 0.97 ( 7.79 / 8) | 45.41 |
| N4374 | 1.88 | 2.00 | 97.97  | 2.18 (−0.22 +0.25) | 39.09 (−10.34 +10.28) | 0.55 ( 2.19 / 4) | 70.06 |
| N4382 | 1.69 | 1.90 | 99.65  | 1.72 (−0.33 +0.37) | 17.48 ( −6.66 + 6.39) | 0.15 ( 0.29 / 2) | 86.36 |
| N4472 | 2.03 | 2.02 | 70.98  | 2.10 (−0.20 +0.22) | 117.08 (−25.06 +25.54) | 0.75 ( 8.23 /11) | 69.23 |
| N4621 | 1.86 | 1.96 | 86.16  | 1.79 (−0.54 +0.72) | 15.30 ( −5.99  +5.57) | 1.63 ( 1.63 / 1) | 20.11 |
| N4636 | 2.10 | 2.04 | 93.78  | 2.10 (−0.35 +0.33) | 46.93 (−14.13 +13.66) | 0.79 ( 3.14 / 4) | 53.44 |
| N4649 | 1.98 | 2.01 | 98.60  | 1.97 (−0.20 +0.22) | 113.52 (−28.32 +30.30) | 0.98 ( 9.80 /10) | 45.79 |
| N4697 | 1.53 | 1.80 |  6.05  | 1.67 (−0.19 +0.20) | 21.55 ( −4.55  +4.55) | 0.92 ( 4.61 / 5) | 46.54 |

col 1: galaxy name
col 2: best fit beta determined by the maximum likelihood method
col 3-4: best fit beta determined by the KS grid search and the corresponding probability
col 5: best fit beta and its error at the 90% confidence level determined by the Chi-square method
col 6: best fit amplitude and its error at the 90% confidence level determined by the Chi-square method
col 7: Chi-square and degrees of freedom determined by the Chi-square method
col 8: probability determined by the Chi-square method

Table 3  Total X-ray Luminostiy of LMXBs

| name (1) | Lx(1) (2) | correction (3) | Lx(2) (4) | L(B) (5) | L(K) (6) | Lx(2)/L(B) (7) | Lx(2)/L(K) (8) |
|---|---|---|---|---|---|---|---|
| I1459 | 2.548 | 0.67 | 3.803 | 6.037 | 30.563 | 0.630 | 0.124 |
| N0720 | 2.354 | 0.68 | 3.462 | 4.269 | 18.583 | 0.811 | 0.186 |
| N1316 | 3.510 | 0.71 | 4.944 | 10.610 | 45.135 | 0.466 | 0.110 |
| N1399 | 6.458 | 0.70 | 9.226 | 4.071 | 22.579 | 2.266 | 0.409 |
| N1407 | 6.828 | 0.79 | 8.643 | 6.742 | 32.719 | 1.282 | 0.264 |
| N3379 | 0.426 | 0.68 | 0.626 | 1.580 | 7.413 | 0.396 | 0.084 |
| N4365 | 2.095 | 0.69 | 3.036 | 2.837 | 11.828 | 1.070 | 0.257 |
| N4374 | 2.021 | 0.67 | 3.016 | 4.415 | 17.951 | 0.683 | 0.168 |
| N4382 | 1.116 | 0.74 | 1.508 | 4.497 | 19.449 | 0.335 | 0.078 |
| N4472 | 6.064 | 0.74 | 8.195 | 8.258 | 38.949 | 0.992 | 0.210 |
| N4621 | 0.917 | 0.68 | 1.349 | 2.735 | 11.130 | 0.493 | 0.121 |
| N4636 | 2.429 | 0.79 | 3.075 | 2.998 | 13.858 | 1.026 | 0.222 |
| N4649 | 6.087 | 0.76 | 8.009 | 5.874 | 29.033 | 1.363 | 0.276 |
| N4697 | 1.445 | 0.80 | 1.806 | 2.081 | 7.755 | 0.868 | 0.233 |

col1: galaxy name
col2: Lx(1) X-ray luminosity (in $10^{40}$ erg/s) of total discrete sources with Lx(min)=$10^{37}$ erg/s and Lx(max)=$2\times10^{39}$ erg/s.
col3: The correction factor is to take into account sources excluded because they are outside S3 and within 20" from the center.
col4: Lx(2) same as Lx(1) after adding sources outside S3 and in the central 20".
col5: L(B)  Optical luminosity in B (in unit of $10^{10}$ solar luminosity; with Mo(B)=5.47 mag).
col6: L(K)  Near-IR luminosity in K (in unit of $10^{10}$ solar luminosity; with Mo(K)=3.33 mag).
col7: Lx(2)/L(B) X-ray to optical luminosity ratio in unit of $10^{30}$ erg/s / Lo
col8: Lx(2)/L(K) X-ray to near-IR luminosity ratio in unit of $10^{30}$ erg/s / Lo

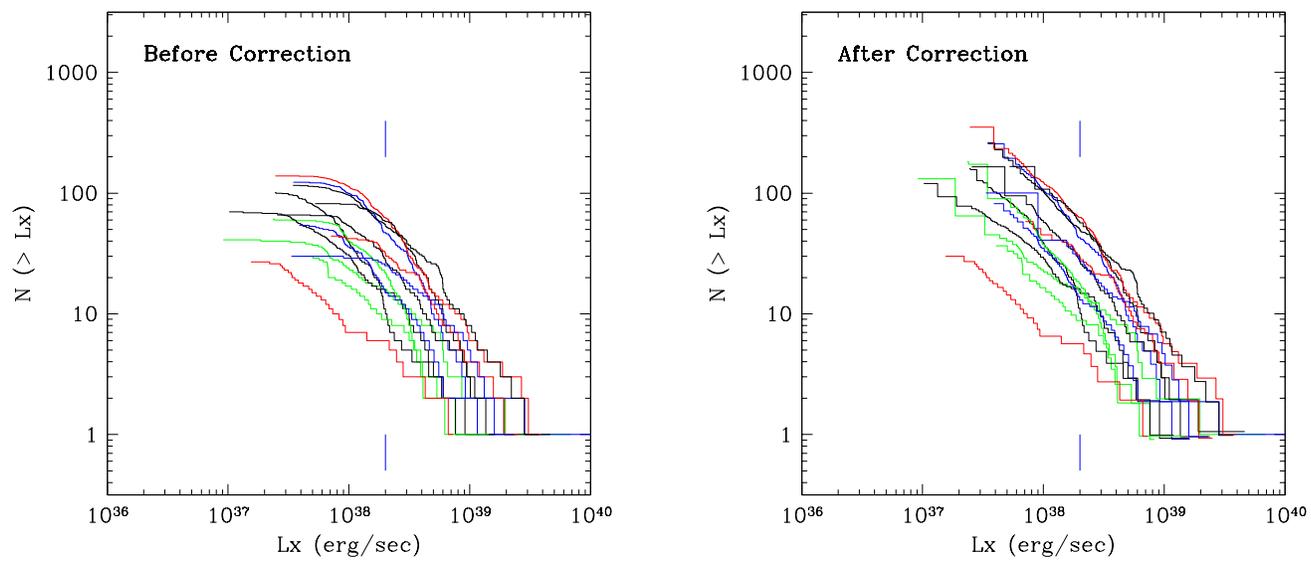

Figure 1:

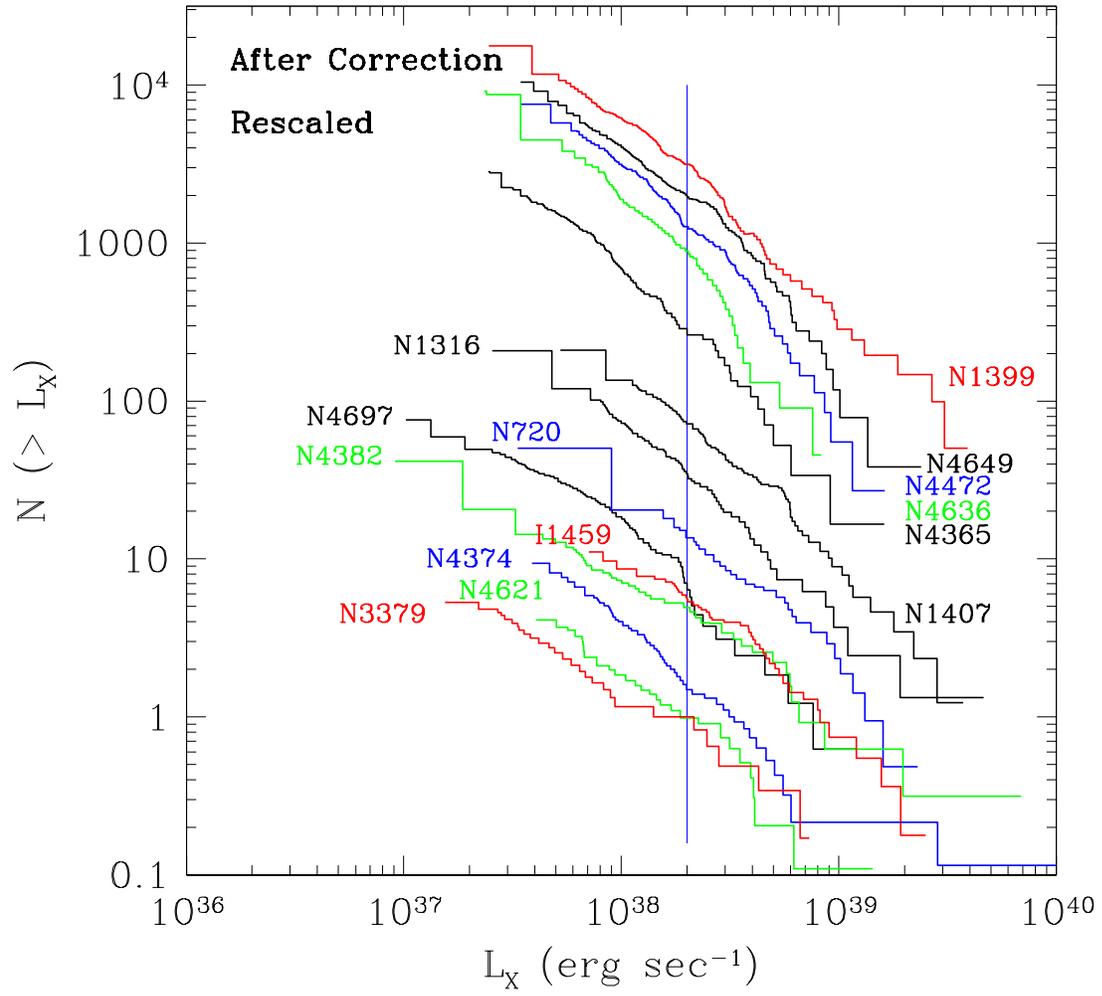

Figure 2:

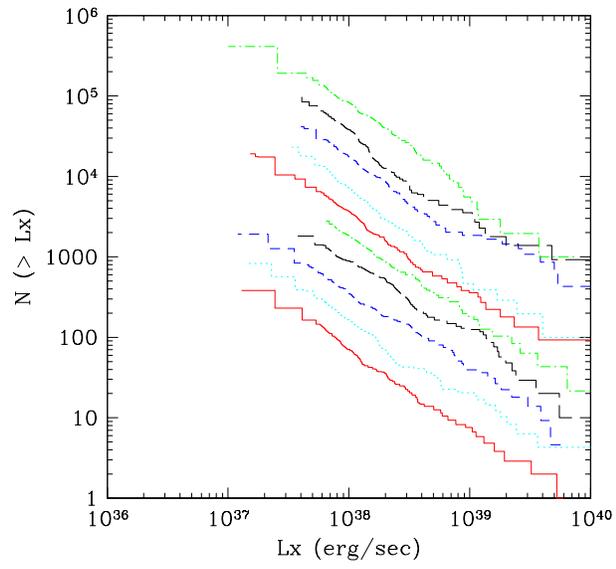

Figure 3:

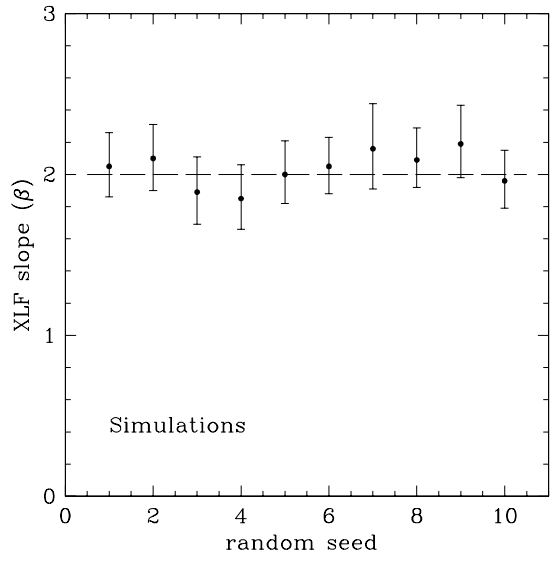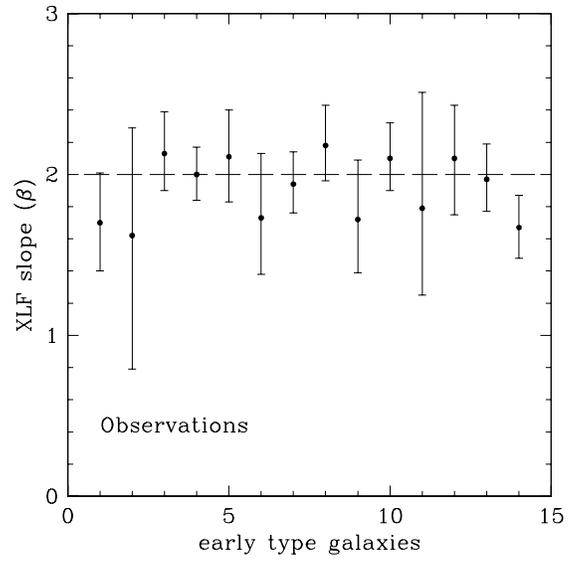

Figure 4:

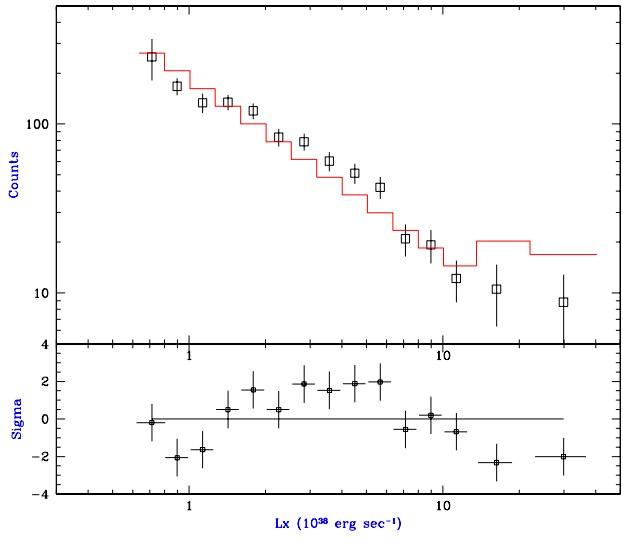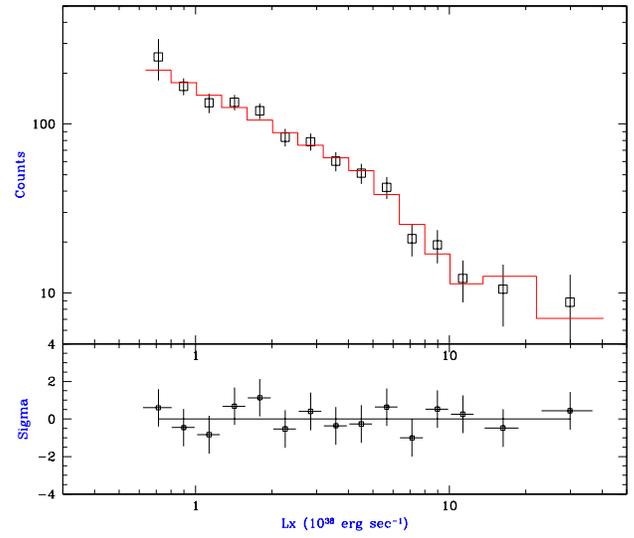

Figure 5:

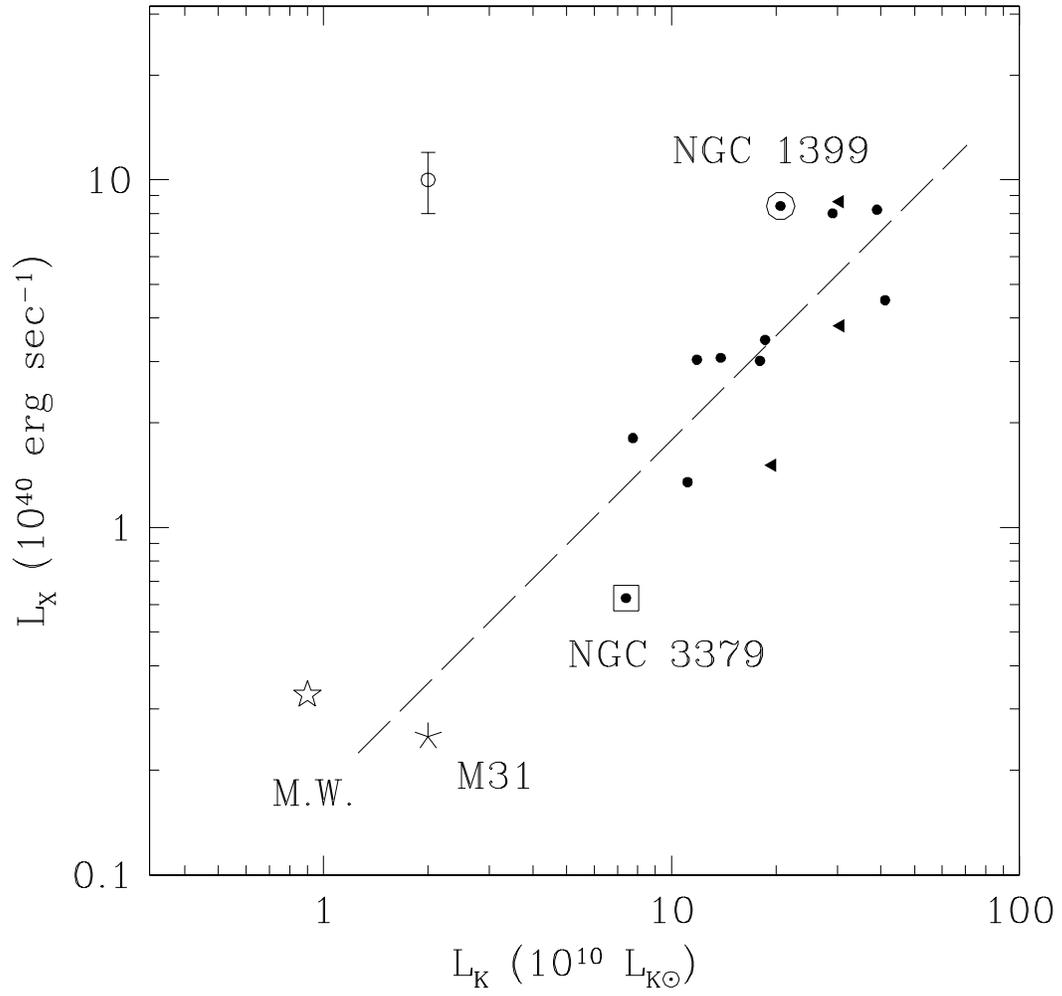

Figure 6:

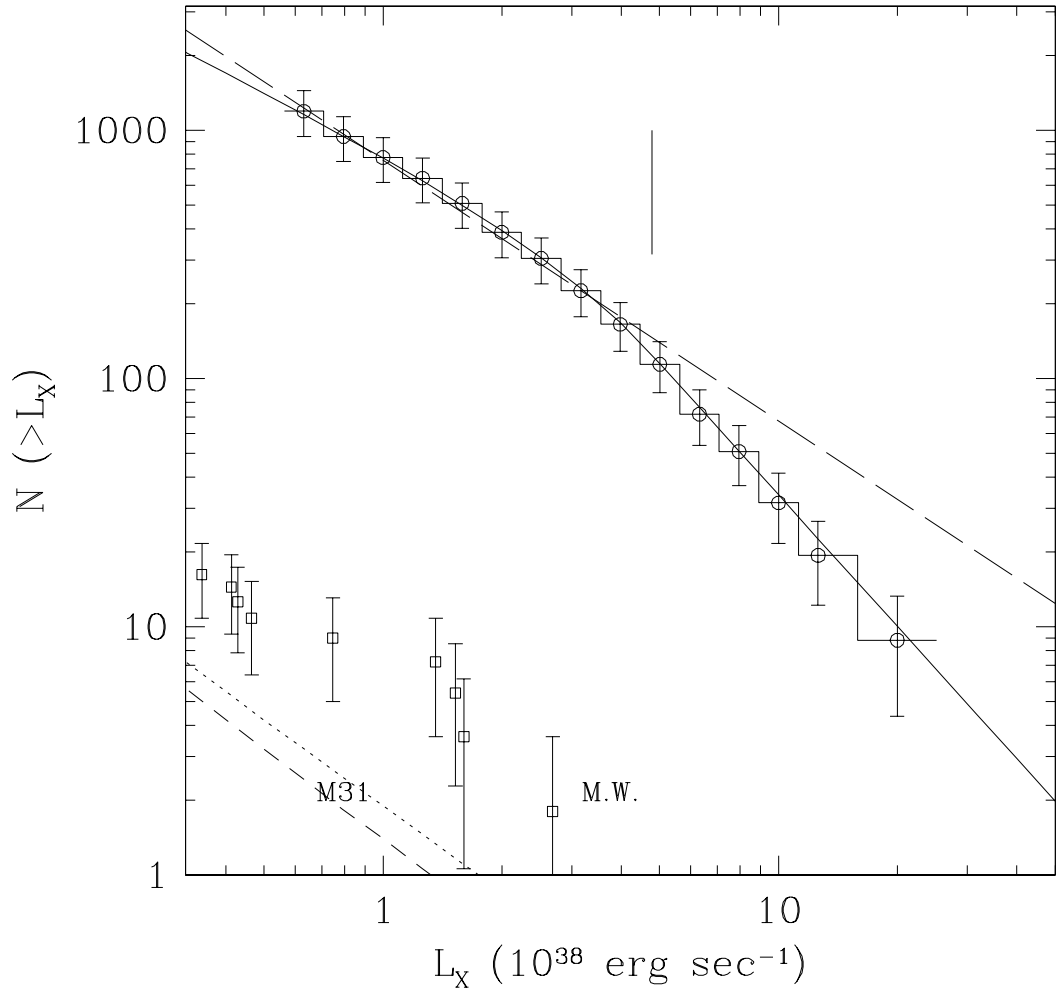

Figure 7:

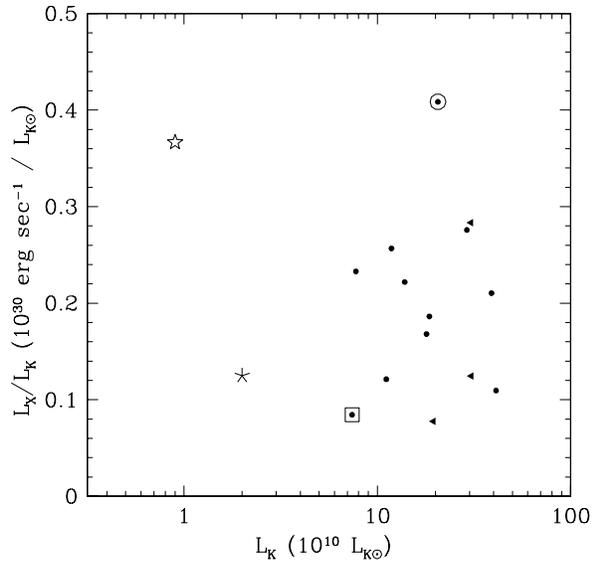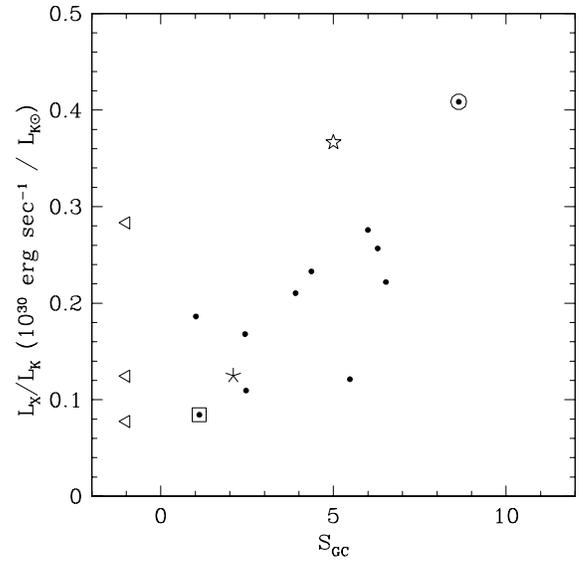

Figure 8: